\documentclass[12p]{iopart}

\usepackage{color}
\usepackage{graphicx}
\usepackage{iopams}
\usepackage{ulem}
\usepackage{braket}
\usepackage[utf8]{inputenc}
\usepackage{url}

\begin{document}

\newcommand{\gtwo}{\(g^{(2)}\)}
\newcommand{\gthree}{\(g^{(3)}\)}
\newcommand{\goneone}{\(g^{(1,1)}\)}

\title{Probing multimode squeezing with correlation functions}
\author{Andreas Christ\(^{1,2}\), Kaisa Laiho\(^2\), Andreas Eckstein\(^2\), Kati\'{u}scia N. Cassemiro\(^2\), and Christine Silberhorn\(^{1,2}\)}
\address{\(^1\)Applied Physics, University of Paderborn, Warburger Straße 100, 33098 Paderborn, Germany}
\address{\(^2\)Max Planck Institute for the Science of Light,\\  G\"unther-Scharowsky Straße 1/Bau 24, 91058 Erlangen, Germany}
\ead{Andreas.Christ@uni-paderborn.de}

\date{\today}

\begin{abstract}
Broadband multimode squeezers constitute a powerful quantum resource with promising potential for different applications in quantum information technologies such as information coding in quantum communication networks or quantum simulations in higher dimensional systems. However, the characterization of a large array of squeezers that coexist in a single spatial mode is challenging. In this paper we address this problem and propose a straightforward method to determine the number of squeezers and their respective squeezing strengths by using broadband multimode correlation function measurements. These measurements employ the large detection windows of state of the art avalanche photodiodes to simultaneously probe the full Hilbert space of the generated state, which enables us to benchmark the squeezed states. Moreover, due to the structure of correlation functions, our measurements are not affected by losses. This is a significant advantage, since detectors with low efficiencies are sufficient. Our approach is less costly than tomographic methods relying on multimode homodyne detection which is based on much more demanding measurement and analysis tools and appear to be impractical for large Hilbert spaces. 
\end{abstract}


\maketitle

\section{Introduction}
The study of correlation functions has a long history and lies at the heart of coherence theory \cite{mandel_optical_1995}. Intensity correlation measurements were first performed by Hanbury Brown and Twiss in the context of classical optics \cite{brown_correlation_1956}. Since then correlation functions have become an standard tool in quantum optical experiments to study the properties of laser beams \cite{chopra_higher-order_1973}, parametric downconversion sources\cite{blauensteiner_photon_2009, ivanova_multiphoton_2006} or heralded single-photons \cite{tapster_photon_1998, uren_characterization_2005, bussieres_fast_2008}. Current state of the art experiments are able to measure correlation functions up to the eighth order \cite{avenhaus_accessing_2010}, giving access to diverse characteristics of photonic states. The normalized second-order correlation function \(g^{(2)}(0)\) probes whether the generated photons are bunched or anti-bunched, with \(g^{(2)}(0) < 1\) being a genuine sign of non-classicality \cite{loudon_quantum_2000}. The measurement of all unnormalized moments \(G^{(n)}\) of a given optical quantum state provide complete access to the photon-number distribution for arbitrary single-mode input states \cite{mandel_optical_1995}. Moreover, it is possible to perform a full state-tomography with the help of correlation function measurements \cite{shchukin_universal_2006}.

The measurement of these correlation functions is, in general, performed in a time resolved manner \(g^{(n)}(t_1, t_2, \dots t_n)\). Limited time resolution has been considered as a detrimental effect and treated as experimental imperfection \cite{tapster_photon_1998}. In contrast to previous work, we employ the finite time resolution of photo-detectors to gain access to the spectral character of broadband multimode quantum states. Our scheme of measuring broadband multimode correlation functions of pulsed quantum light is especially useful for probing squeezed states. These states are commonly generated via the interaction of light with a crystal exhibiting a \(\chi^{(2)}\)-nonlinearity, a process referred to as parametric downconversion (PDC)\cite{rarity_quantum_1992, mauerer_how_2009, wasilewski_pulsed_2006, lvovsky_decomposing_2007, wenger_pulsed_2004, anderson_pulsed_1997} or with optical fibers featuring a \(\chi^{(3)}\)-nonlinearity called four-wave-mixing (FWM) \cite{loudon_squeezed_1987, levenson_generation_1985}. 

In general the generated squeezed states exhibit multimode characteristics in the spectral degree of freedom, i.e. a set of independent squeezed states is created with each squeezer residing in its own Hilbert space. This inherent multimode character renders these states powerful for coding quantum information, yet the same feature impedes a proper experimental characterization in a straightforward manner. Due to the sheer vastness of the corresponding Hilbert space, standard quantum tomography methods become time-consuming and ineffective. It is neither easy to determine the degree of squeezing in each mode, nor the amount of generated independent squeezers. Nonetheless, these are the key benchmarks defining the potential of a source for quantum information and quantum cryptography applications. In the following we investigate how to overcome these issues and elaborate on an alternative approach to determine the properties of multimode squeezed states based on measuring broadband multimode correlation functions. 

This paper is structured as follows: In section \ref{sec:multimode_squeezer} we revisit the general structure of multimode twin-beam squeezers drawing special attention --- but not restricting --- to states generated by parametric downconversion and four-wave-mixing. Section \ref{sec:correlation_functions} presents the formalism of correlation functions, introduces the intricacies of finite time resolution and defines broadband multimode correlation measurements. Section \ref{sec:probing_twin_beam_squeezing} combines the findings of sections \ref{sec:multimode_squeezer} and \ref{sec:correlation_functions}: We analyze the relation between the number of generated squeezers, their respective squeezing strengths and broadband multimode correlation functions, which leads us to proposing our scheme for characterizing multimode squeezing with the aid of broadband multimode correlation functions.

\section{Multimode Squeezers}\label{sec:multimode_squeezer}
In a squeezed state of light one quadrature of the field exhibits an uncertainty below the standard quantum level at the expense of an increased variance in the conjugate quadrature, such that the Heisenberg's uncertainty relation holds at its minimum attainable value. The standard description of squeezed states usually considers two different types of squeezers: single-beam squeezers and twin-beam squeezers. Single-beam squeezers create the squeezing into a single optical mode \(\hat{S} = \exp\left(-\zeta \hat{a}^{\dagger2} + \zeta^* \hat{a}^2 \right)\), whereas twin-beam squeezers consist of \textit{two} beams with inter-beam squeezing \(\hat{S}^{ab} = \exp\left(-\zeta \hat{a}^\dagger \hat{b}^\dagger + \zeta^* \hat{a} \hat{b}\right) \) \cite{barnett_methods_2003}. In these equations \(\zeta\) labels the squeezing strength and the operators \(\hat{a}^\dagger, \hat{b}^\dagger\) create photons in distinct optical modes.

In this section we go beyond the standard description and discuss the theory of squeezed states, which are generated by the interaction of ultrafast pump pulses with nonlinear crystals or optical fibers. Here, we concentrate on the spectral structure of the broadband output beams. In general the utilized optical processes, typically called optical parametric amplification (OPA) or parametric downconversion (PDC) do not generate one but a variety of different squeezers in multiple frequency modes. A whole set of independent squeezed beams is generated in broadband orthogonal spectral modes within an optical beam. We refer to these states as frequency multimode single- or twin-beam squeezers \cite{wasilewski_pulsed_2006}. Here the \textit{multimode} prefix indicates that more than one squeezer is present in the optical beam and the term \textit{single- or twin-beam} identifies whether one squeezed beam or two entangled squeezed beams are created. Due to the single-pass configuration of our sources losses are negligible, hence we restrict ourselves to the analysis of pure squeezed states. 

\subsection{Multimode twin-beam squeezers}
The subject of our analysis is twin-beam squeezing generated by the propagation of an ultrafast pump pulse through a nonlinear medium (single-beam squeezers are discussed in \ref{app:single_beam_squeezer}). For simplicity we focus on the collinear propagation of all involved fields each generated into a single spatial mode. This description is rigorously fulfilled for PDC in waveguides \cite{mosley_direct_2009, christ_spatial_2009}, but can also be applied for other experimental configurations, since the approximation carries all the complexities of the multimode propagation in the spectral degree of freedom. If the pump field is undepleted, we can neglect its quantum fluctuations and describe this OPA process by the effective quadratic Hamiltonian
(see \ref{app:multimode_two_mode_squeezer_generation} for a detailed derivation)
\begin{eqnarray}
    \hat{H}_{OPA} = A \int \mathrm d \omega_s \int \mathrm d \omega_i\, f(\omega_s, \omega_i) \hat{a}_s^\dagger(\omega_s) \hat{a}_i^\dagger(\omega_i) + h.c. \, ,
    \label{eq:effective_OPA_hamiltonian_two_mode}
\end{eqnarray}
in which the constant \(A\) denotes the overall efficiency of the OPA, the function \(f(\omega_s, \omega_i)\) describes the normalized output spectrum of the downconverted beam, which --- in many cases --- is close to a two-dimensional Gaussian distribution. The operators \(\hat{a}^\dagger_s(\omega_s)\) and \(\hat{a}^\dagger_i(\omega_i)\) are the photon creation operators in the different twin-beam arms, in general labelled signal and idler, respectively.

The unitary transformation generated by the effective OPA Hamiltonian in equation \eref{eq:effective_OPA_hamiltonian_two_mode} can be written in the form  
\begin{eqnarray}
    \fl \qquad \hat{U}_{OPA} &= \exp\left[-\frac{\imath}{\hbar}  \left( A \int \mathrm d \omega_s \int \mathrm d \omega_i\, f(\omega_s, \omega_i) \hat{a}_s^\dagger(\omega_s) \hat{a}_i^\dagger(\omega_i) + h.c. \right) \right].
    \label{eq:effective_OPA_unitary_two_mode_hamilton}
\end{eqnarray}
By virtue of the singular-value-decomposition theorem \cite{law_continuous_2000} we decompose the two terms in the exponential of equation \eref{eq:effective_OPA_unitary_two_mode_hamilton} as
\begin{eqnarray}
    \nonumber
     -\frac{\imath}{\hbar} A f(\omega_s, \omega_i) =  \sum_k r_k \psi^*_k(\omega_s) \phi^*_k(\omega_i), \,\,\,\, \mathrm{and}\\
    -\frac{\imath}{\hbar} A^* f^*(\omega_s, \omega_i) = - \sum_k r_k \psi_k(\omega_s) \phi_k(\omega_i).
    \label{eq:singular_value_decomposition}
\end{eqnarray}
Here both \(\left\{\psi_k(\omega_s)\right\}\) and \(\left\{\phi_k(\omega_i)\right\}\) each form a complete set of orthonormal functions. The amplitudes of the generated modes \(\psi_k(\omega_s)\) and \(\phi_k(\omega_i)\) are given by the \(r_k \in \mathbb{R}^+\) distribution.
Employing equation \eref{eq:singular_value_decomposition} and introducing a new broadband mode basis \cite{rohde_spectral_2007} for the generated state as:
\begin{eqnarray}
    \hat{A}_k = \int \mathrm d \omega_s \psi_k(\omega_s) \hat{a}_s(\omega_s)  \,\,\,\, \mathrm{and} \,\,\,\,
    \hat{B}_k = \int \mathrm d \omega_i \phi_k(\omega_i) \hat{a}_i(\omega_i),
\end{eqnarray}
we obtain the unitary transformation \cite{mauerer_how_2009}
\begin{eqnarray}
    \nonumber
    \hat{U}_{OPA} &= \exp\left[\sum_k r_k \hat{A}_k^\dagger \hat{B}_k^\dagger - h.c. \right] \\
    \nonumber
                  &= \bigotimes_k \exp\left[r_k \hat{A}_k^\dagger \hat{B}_k^\dagger - h.c. \right] \\
                  &= \bigotimes_k \hat{S}^{ab}_k(-r_k).
    \label{eq:effective_OPA_unitary_two_mode}
\end{eqnarray}
In total the OPA generates a tensor product of distinct broadband twin-beam squeezers as defined in \cite{barnett_methods_2003} with squeezing amplitudes \(r_k\), related to the available amount of squeezing via: \(\mathrm{squeezing[dB]} = -10 \log_{10}\left(e^{-2 r_k}\right)\). The Heisenberg representation of the multimode twin-beam squeezers is given by independent input-output relations for each broadband beam 
\begin{eqnarray}
    \nonumber
        \hat{A}_k \Rightarrow \cosh(r_k) \hat{A}_k + \sinh(r_k)\hat{B}_k^\dagger \\
        \hat{B}_k \Rightarrow \cosh(r_k) \hat{B}_k + \sinh(r_k)\hat{A}_k^\dagger .
    \label{eq:two_mode_squeezer_input_output_relation}
\end{eqnarray}
Note that the squeezer distribution \(r_k\) and basis modes \(\hat{A}_k\) and \(\hat{B}_k\) are unique and well-defined properties of the generated twin-beam. Their exact form is given by the Schmidt decomposition of the joint spectral amplitude \(-\frac{\imath}{\hbar} A f(\omega_s, \omega_i)\). This mathematical transformation directly yields the physical shape of the generated optical modes \(\psi_k(\omega_s)\), \(\phi_k(\omega_i)\) with each pair \(\hat{A}_k\) and  \(\hat{B}_k\) being strictly correlated.

\begin{figure}[htpb]
    \begin{center}
        \includegraphics[width=0.8\linewidth]{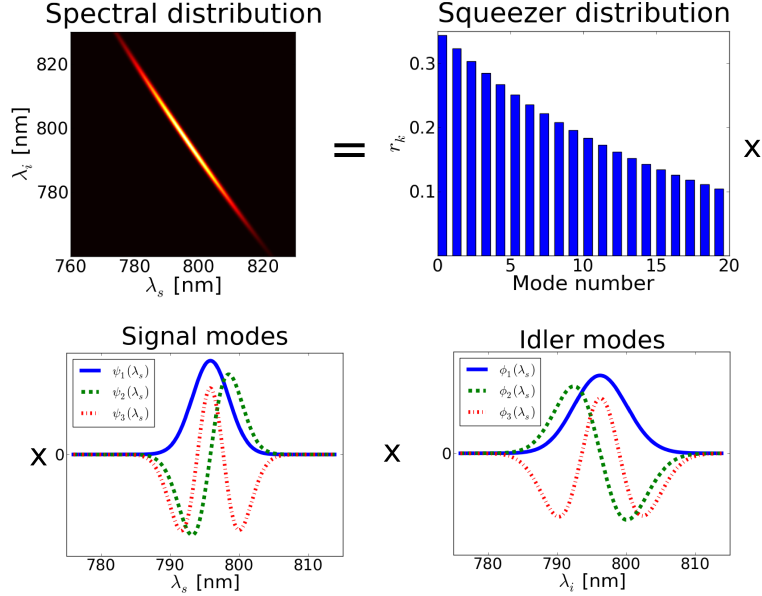}
    \end{center}
    \caption{Visualization of the singular value decomposition in equation \eref{eq:singular_value_decomposition}. The frequency distribution \(- \frac{\imath}{\hbar} A f(\omega_s, \omega_i)\) of the generated state defines the shape of the signal and idler modes \(\psi_k(\omega_s), \phi_k(\omega_i)\) and the squeezer distribution \(r_k\).}
    \label{fig:schmidt_modes}
\end{figure}

In figure \ref{fig:schmidt_modes} we illustrated one possible squeezer distribution and corresponding broadband modes. The joint spectral distribution \(f(\omega_s, \omega_i)\) of the generated twin-beams shown in figure \ref{fig:schmidt_modes} defines the shape of the broadband signal and idler modes \(\hat{A}_k\) and \(\hat{B}_k\). In the special case of a Gaussian spectral distribution the form of the squeezing modes resembles the Hermite functions. The number of different squeezer modes is closely connected to the frequency correlations between the signal and idler beam. In the presented case the spectrally correlated beams lead to over 20 independent squeezers. The total amount of squeezing depends on the constant \(A\) appearing in the Hamiltonian in equation \eref{eq:effective_OPA_hamiltonian_two_mode}, which is directly related to the applied pump power \(I\) and the strength of the nonlinearity  \(\chi^{(2)}\) in the medium \( (A \propto \sqrt{I}, \chi^{(2)})\).

The OPA state is mainly characterized by the number of squeezed modes and the overall gain of the process, both being determined by the distribution of the individual squeezing amplitudes \(r_k\). In order to analyze the number of generated squeezers independently from the amount of squeezing, we split the distribution of squeezing weights \(r_k\) into a normalized distribution \(\lambda_k\)  \(\left(\sum_k \lambda_k^2 = 1\right)\) that characterizes the probability for occupation of different squeezers in the respective optical quantum state, and an overall gain of the process \(B \in \mathbb{R}^+\), quantifying the total amount of generated squeezing according to
\begin{eqnarray}
    r_k = B \, \lambda_k.
\end{eqnarray}
The characterization of these two fundamental properties of a multimode twin-beam state is a major experimental challenge. While these states are easily generated in the lab, a tomography by means of homodyne detection would require to match for each squeezed mode \(\hat{A}_k\) and \(\hat{B}_k\) different local oscillator beams with adapted temporal-spectral pulse shapes. Multimode homodyning \cite{beck_joint_2001} may provide a route to circumvent this difficulty, however an experimental implementation still appears challenging.

\section{Correlation functions}\label{sec:correlation_functions}
The n-th order (normalized) correlation function \(g^{(n)}(t_1, t_1, \dots ,t_n)\) is generally defined as a time-dependent function of the electromagnetic field. For quantized electric field operators, it can be expressed as \cite{glauber_quantum_1963, loudon_quantum_2000, mandel_optical_1995, vogel_quantum_2006} 
\begin{eqnarray}
     g^{(n)}(t_1, t_2, \dots, t_n) 
    =\frac{\left< \hat{E}^{(-)}(t_1) \dots \hat{E}^{(-)}(t_n)\hat{E}^{(+)}(t_1) \dots \hat{E}^{(+)}(t_n)\right>}
    {\left< \hat{E}^{(-)}(t_1)\hat{E}^{(+)}(t_1)\right> \dots \left< \hat{E}^{(-)}(t_n) \hat{E}^{(+)}(t_n) \right>},
    \label{eq:correlation_function-time_resolved}
\end{eqnarray}
and it measures the (normalized) n-th order temporal correlations at different points in time. Note that this definition of the correlation functions is independent of coupling losses and detection inefficiencies yielding a loss resilient measure \cite{avenhaus_accessing_2010}. Realistic detectors however, suffer from internal jitter and finite gating times. We accommodate for these resolution effects by weighting the correlation function with the appropriate detection window \(T(t)\) of the applied detectors as presented in \cite{tapster_photon_1998}, and obtain
\begin{figure}[htpb]
    \begin{center}
        \includegraphics[width=0.9\linewidth]{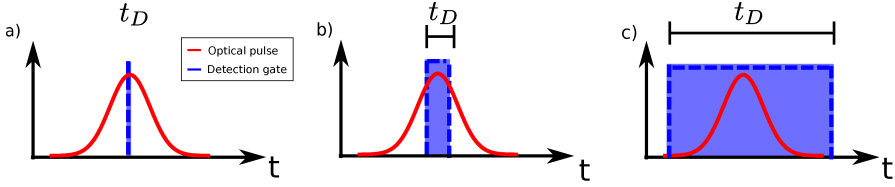}
    \end{center}
    \caption{a) perfect time-resolved detection; b) finite detection gate; c) broadband detection gate exceeding the pulse duration giving rise to different types of correlation measures.}
\label{fig:correlation_function}
\end{figure}

\begin{eqnarray}
    \nonumber
    \fl g^{(n)}(t_1, t_2, \dots, t_n) = \\
    \fl \qquad \frac{\int \mathrm d t_1 T(t_1) \dots \int \mathrm d t_n T(t_n) \left< \hat{E}^{(-)}(t_1) \dots \hat{E}^{(-)}(t_n)\hat{E}^{(+)}(t_1) \dots \hat{E}^{(+)}(t_n)\right>}
    { \int  \mathrm d t_1 T(t_1) \left< \hat{E}^{(-)}(t_1)\hat{E}^{(+)}(t_1)\right> \dots \int  \mathrm d t_n T(t_n) \left< \hat{E}^{(-)}(t_n) \hat{E}^{(+)}(t_n) \right>}.
    \label{eq:correlation_function-time_resolved_finite_gating}
\end{eqnarray}
If the employed photo-detectors exhibit flat detection windows, exceeding the length of the investigated pulses (\(T(t) \rightarrow \mathrm{const.}\)), equation \eref{eq:correlation_function-time_resolved_finite_gating} can be simplified to
\begin{eqnarray}
    g^{(n)} = 
    \frac{\int \mathrm d t_1  \dots \mathrm d t_n \left< \hat{E}^{(-)}(t_1) \dots \hat{E}^{(-)}(t_n)\hat{E}^{(+)}(t_1) \dots \hat{E}^{(+)}(t_n)\right>}
    { \int  \mathrm d t_1 \left< \hat{E}^{(-)}(t_1)\hat{E}^{(+)}(t_1)\right> \dots \int  \mathrm d t_n \left< \hat{E}^{(-)}(t_n) \hat{E}^{(+)}(t_n) \right>}.
    \label{eq:correlation_function-time_resolved_flat}
\end{eqnarray}
This theoretical model is adequate for the detection of ultrafast pulses with standard avalanche photodetectors. Furthermore, equation \eref{eq:correlation_function-time_resolved_flat} exhibits the convenient property of time independence and represents our generalized broadband multimode correlation function. Despite its similarity to the common correlation functions as defined in  equation \eref{eq:correlation_function-time_resolved}, the broadband multimode correlation function in  equation \eref{eq:correlation_function-time_resolved_flat} should no longer be considered as a naive general measure of n-th order coherence. In figure \ref{fig:correlation_function} we illustrate the main difference between the time-integrated and time-resolved correlation measurements. 

Equation \eref{eq:correlation_function-time_resolved_flat} is still not optimal for our studies of squeezed light fields. We transform it further by replacing the electric field operators by photon number creation and destruction operators (\(\hat{E}^{(+)}(t_n) \propto \hat{a}(t_n)\)) and perform a Fourier transform from the time domain into the frequency domain (\( \hat{a}(t) = \int \mathrm d \omega\, \hat{a}(\omega) e^{-\imath \omega t}\)). Equation \eref{eq:correlation_function-time_resolved_flat} is then rewritten as 
\begin{eqnarray}
    \nonumber
    g^{(n)} &=
     \frac{\int \mathrm d \omega_1  \dots \mathrm d \omega_n \left< \hat{a}^\dagger(\omega_1) \dots \hat{a}^\dagger(\omega_n)\hat{a}(\omega_1) \dots \hat{a}(\omega_n)\right>}
    { \int  \mathrm d \omega_1 \left< \hat{a}^\dagger(\omega_1)\hat{a}(\omega_1)\right> \dots \int  \mathrm d \omega_n \left< \hat{a}^\dagger(\omega_n) \hat{a}(\omega_n) \right>} \\
    &= \frac{\left<: \left( \int \mathrm d \omega \hat{a}^\dagger(\omega) \hat{a}(\omega) \right)^n: \right>}{\left< \int \mathrm d \omega \hat{a}^\dagger(\omega) \hat{a}(\omega) \right>^n},
    \label{eq:correlation_function-time_resolved_flat_frequency_domain}
\end{eqnarray}
in which \(\left<: \cdots :\right>\) indicates normal ordering of the enclosed photon creation and destruction operators. In addition we adapt the correlation function to the basis of the measured quantum system, i.e. we perform a general basis transform from \(\hat{a}(\omega)\) to the basis of the measured multimode twin-beam squeezers \(\hat{A}_k\). This results in:
\begin{eqnarray}
    g^{(n)} = \frac{\left<:\left(\sum_k \hat{A}_k^\dagger\hat{A}_k\right)^n:\right>}{\left<\sum_k \hat{A}_k^\dagger\hat{A}_k\right>^n}
\label{eq:broadband_correlation_function_final}
\end{eqnarray}
Equations \eref{eq:correlation_function-time_resolved_flat}, \eref{eq:correlation_function-time_resolved_flat_frequency_domain} and \eref{eq:broadband_correlation_function_final} stress the key difference between time-resolved and time-integrated correlation function measurements. While time-resolved correlation functions probe specific temporal modes, time-integrating detectors directly measure a superposition of all the different modes. This specific feature of broadband multimode detection is essential for our analysis. The simultaneous measurement of all different optical modes gives us direct \textit{loss-independent} access to the squeezer distribution of the probed state.

\subsection{Broadband multimode cross-correlation functions}
In the previous section we restricted ourselves to intra-beam correlations. To allow for measurements of correlations between different beams we extend our analysis. The identification of such inter-beam correlations is of special importance in quantum optics and quantum information applications, since they quantify the continuous variable entanglement between different subsystems, in our case the analyzed optical beams. In section \ref{sec:multimode_squeezer} we have already discussed one of the most employed entanglement sources: Twin-beam squeezers. These states are not only entangled in their quadratures, but also in their spectral and spatial degrees of freedom \cite{braunstein_quantum_2005}. In order to probe higher-order cross-correlations between the two different beams \cite{vogel_quantum_2006}, or subsystems \(a\) and \(b\) of order \(n\) and \(m\) respectively, we generalize equation \eref{eq:correlation_function-time_resolved} to
\begin{eqnarray}
    \nonumber
    & \fl  g^{(n, m)}(t^{(a)}_1, t^{(a)}_2, \dots, t^{(a)}_n; t^{(b)}_1, t^{(b)}_2, \dots, t^{(b)}_m) =\\
    & \fl =\frac{\left< \hat{E}_a^{(-)}(t^{(a)}_1) \dots \hat{E}_a^{(-)}(t^{(a)}_n)\hat{E}_a^{(+)}(t^{(a)}_1) \dots \hat{E}_a^{(+)}(t^{(a)}_n) \times \hat{E}_b^{(-)}(t^{(b)}_1) \dots  \hat{E}_b^{(+)}(t^{(b)}_n)\right>}
    {\left< \hat{E}_a^{(-)}(t^{(a)}_1)\hat{E}_a^{(+)}(t^{(a)}_1)\right> \dots \left< \hat{E}_a^{(-)}(t^{(a)}_n) \hat{E}_a^{(+)}(t^{(a)}_n) \right> \times \dots \left<\hat{E}_b^{(-)}(t^{(b)}_n)\hat{E}_b^{(+)}(t^{(b)}_n)\right>}.
    \label{eq:cross_correlation_function-time_resolved}
\end{eqnarray}
Taking into account broadband detection windows --- exceeding the pulse duration --- the above formula can be reformulated as 
\begin{eqnarray}
    g^{(n,m)} = \frac{\left<:\left(\int\mathrm d t\,\hat{E}_a^{(-)}(t)\hat{E}_a^{(+)}(t)\right)^n: :\left(\int\mathrm d t \,\hat{E}_b^{(-)}(t)\hat{E}_b^{(+)}(t)\right)^m: \right>}{\left<\int\mathrm d t \,\hat{E}_a^{(-)}(t)\hat{E}_a^{(+)}(t)\right>^n\left<\int\mathrm d t \,\hat{E}_b^{(-)}(t)\hat{E}_b^{(+)}(t)\right>^m}. 
\end{eqnarray}
Again we perform the same simplifications as in equation \eref{eq:correlation_function-time_resolved_flat_frequency_domain} in section \ref{sec:correlation_functions}, namely we replace the  electric field operators by photon creation and destruction operators, apply the Fourier transform from the time to frequency domain and finally we adapt the measurement basis to the given optical state. We find an extended version of equations \eref{eq:correlation_function-time_resolved_flat_frequency_domain} and  \eref{eq:broadband_correlation_function_final}
\begin{eqnarray}
    g^{(n,m)} &= \frac{\left<:\left(\int\mathrm d\omega \,\hat{a}^\dagger(\omega)\hat{a}(\omega)\right)^n: :\left(\int\mathrm d\omega \,\hat{b}^\dagger(\omega)\hat{b}(\omega)\right)^m: \right>}{\left<\int\mathrm d\omega \,\hat{a}^\dagger(\omega)\hat{a}(\omega)\right>^n\left<\int\mathrm d\omega \,\hat{b}^\dagger(\omega)\hat{b}(\omega)\right>^m} \\
     &= \frac{\left<:\left(\sum_k \hat{A}_k^\dagger\hat{A}_k\right)^n::\left(\sum_k \hat{B}_k^\dagger\hat{B}_k\right)^m:\right>}{\left<\sum_k \hat{A}_k^\dagger\hat{A}_k\right>^n\left<\sum_k \hat{B}_k^\dagger\hat{B}_k\right>^m}.
\label{eq:broadband_cross-correlation_function}
\end{eqnarray}
Further extensions of cross-correlation measurements to systems consisting of more than two different beams are possible \cite{mandel_optical_1995}, but are not necessary within the scope of this paper.

\section{Probing frequency multimode squeezers via correlation functions}\label{sec:probing_twin_beam_squeezing}
Using the theoretical description of squeezers as well as the derived broadband multimode correlation functions, we now combine the findings of section \ref{sec:multimode_squeezer} and \ref{sec:correlation_functions}. We establish a connection between the broadband multimode correlation functions and the properties of the squeezing, i.e. the mode distribution \(\lambda_k\) and the optical gain \(B\). 

\subsection{Probing the number of modes via \gtwo-measurements}\label{sec:probing_twin_beam_squeezing_mode_distribution}
\begin{figure}[htpb]
    \begin{center}
        \includegraphics[width=0.91\linewidth]{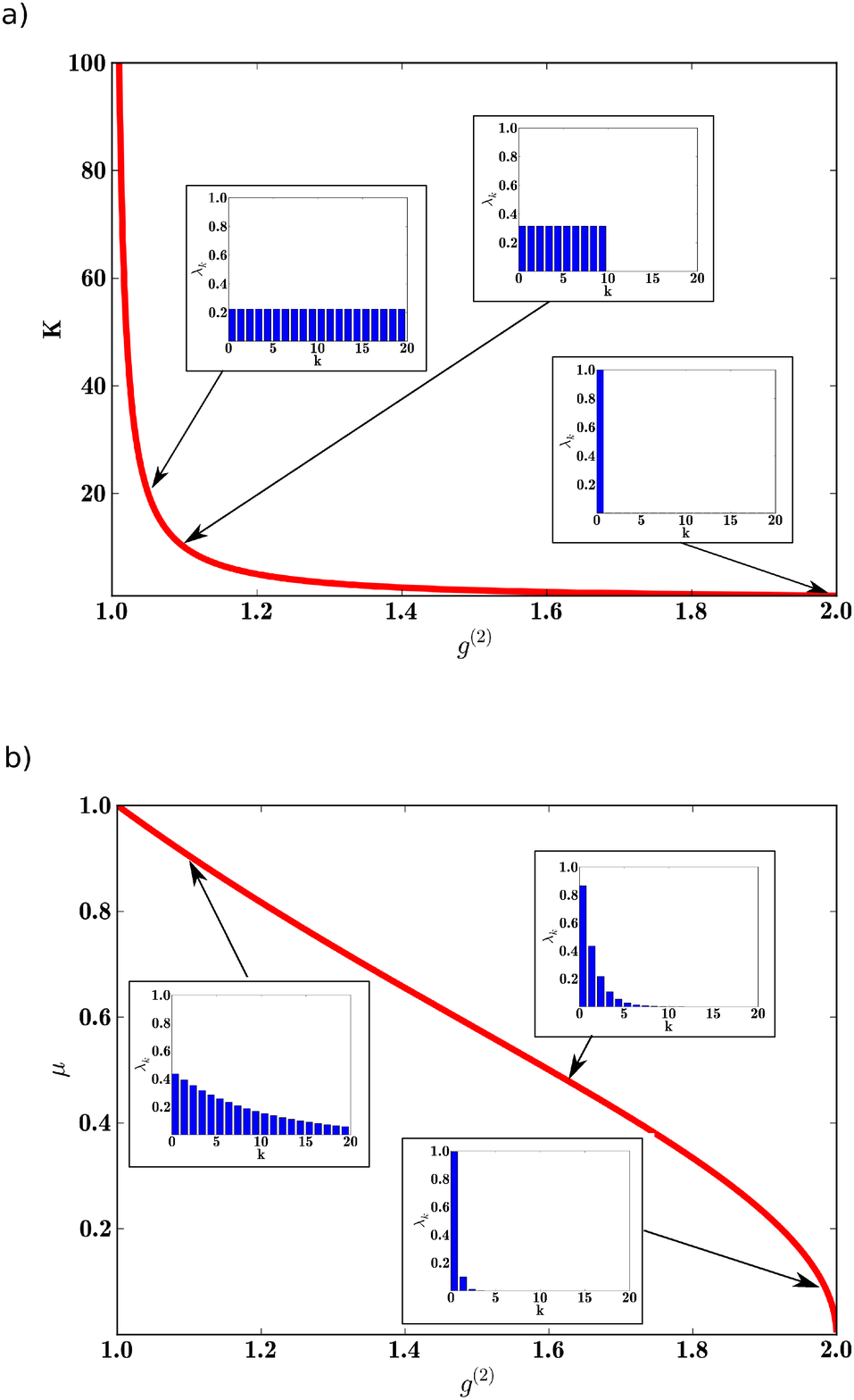}
    \end{center}
    \caption{a) Plot of the effective mode number \(K\) as a function of \(g^{(2)}\) for various effective numbers of modes. b) Visualization of \(\mu\) as a function of \(g^{(2)}\) for different thermal squeezer distributions.}
    \label{fig:g2_mm_squeezer_results}
\end{figure}

The foremost important property of frequency multimode squeezers is the number of independent squeezers in the generated twin-beam state, which is specified by the mode distribution \(\lambda_k\). In contrast to the optical gain \(B\), which is easily tuned by adjusting the pump power the mode distribution \(\lambda_k\) is heavily constricted by the dispersion in the nonlinear material and hence --- in general --- not easily adjustable \cite{frequency_filter}. The effective number of modes in multimode twin-beam state is given by the Schmidt number or cooperativity parameter \(K\) as defined in \cite{eberly_schmidt_2006, r_grobe_measure_1994} with
\begin{eqnarray}
    K = 1 / \sum_k \lambda_k^4.
    \label{eq:schmidt_number}
\end{eqnarray}
Under the assumption of an independent uniform squeezer distribution it directly reflects the number of occupied modes.  
\begin{figure}[htpb]
    \begin{center}
        \includegraphics[width=0.6\linewidth]{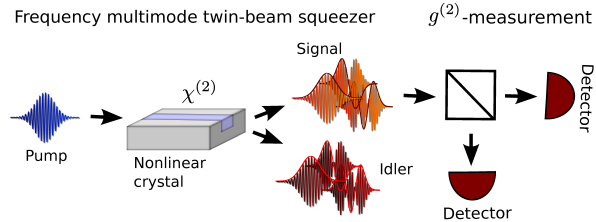}
    \end{center}
    \caption{Setup to measure \(g^{(2)}\) of a multimode twin-beam squeezer.}
    \label{fig:mm_two_mode_squeezr_g2_setup}
\end{figure}
The mode number \(K\) of a multimode twin-beam squeezer can be directly accessed by measuring the broadband multimode  \(g^{(2)}\)-correlation function in the signal or idler arm as depicted in figure \ref{fig:mm_two_mode_squeezr_g2_setup}. This is a result of the structure of the second-order correlation function, which --- by using  \eref{eq:broadband_correlation_function_final} and \eref{eq:two_mode_squeezer_input_output_relation} --- can be expressed as
\begin{eqnarray}
        g^{(2)} &= 1 + \frac{\sum_k \sinh^4(r_k) } { \left[\sum_k \sinh^2(r_k)\right]^2 }.
        \label{eq:gtwo_mm_two_mode_squeezer_nonlinear}
\end{eqnarray}
For our further analysis it is useful to distinguish the low gain from the high gain regime, corresponding to low and high levels of squeezing. In the low gain regime corresponding to biphotonic states typically referred to in the context of PDC experiments \(\sinh(r_k) \approx r_k = B \lambda_k\) and we are able to simplify equation \eref{eq:gtwo_mm_two_mode_squeezer_nonlinear} to
\begin{eqnarray}
    \nonumber
    g^{(2)} &\approx 1 + \frac{\sum_k B^4 \lambda_k^4) } { \left(\sum_k B^2 \lambda_k^2\right)^2 }
    = 1 + \frac{\sum_k  \lambda_k^4 } { \left(\sum_k  \lambda_k^2\right)^2 }
    = 1 + \sum_k  \lambda_k^4\\
    &=1 + \frac{1}{K}.
    \label{eq:gtwo_mm_two_mode_squeezer}
\end{eqnarray}
Consequently the effective number of modes is directly available from the correlation function measurement via \(K = 1 / ( g^{(2)} -1 )\). For a single twin-beam squeezer (\(K = 1\)) \(g^{(2)} = 2\), whereas for higher numbers of squeezers (\(K \gg 1\)) the contributions from the term \(\sum_k \lambda_k^4\) becomes negligible and \(g^{(2)}\) approaches one. This direct correspondence between \(g^{(2)}\) and the effective number of modes \(K\) is presented in figure \ref{fig:g2_mm_squeezer_results} (a). 

Another way of interpreting equation \eref{eq:gtwo_mm_two_mode_squeezer} is to approach the correlation function measurement from the  photon-number point of view. The \(g^{(2)}\)-value of a single twin-beam squeezer, which exhibits a thermal photon-number distribution, evaluates to \gtwo\(=2\). If more squeezers are involved the detector cannot distinguish between the different thermal distributions, i.e. it measures a convolution of all the different thermal photon streams, which gives a Poissonian photon-number distribution \cite{mauerer_how_2009, avenhaus_photon_2008}. In fact one can show that the \gtwo-correlation function in equation \eref{eq:gtwo_mm_two_mode_squeezer_nonlinear} is the convolution of the second-order moments of each individual squeezer. 

Once more, we stress that the \gtwo-measurement does not give access to the exact distribution of squeezers \(\lambda_k\), but to the \textit{effective} number of modes under the assumption that all squeezed states share an identical amount of squeezing. This is a rather crude model and does not fit very well to many experimental realizations. Fortunately, there is a common class of squeezed states, for which a much more refined mode distribution \(\lambda_k\) is accessible: In the case of a two-dimensional Gaussian joint-spectral distribution \(f(\omega_s, \omega_i)\), the distribution \(\lambda_k\) is thermal \(\lambda_k = \sqrt{1 - \mu^2} \mu^k\), and thus it can be characterized by a single distribution parameter \(\mu\) \cite{uren_photon_2003}. The latter can be retrieved from a \gtwo-measurement via \(\mu = \sqrt{2 / g^{(2)} - 1}\), as depicted in figure \ref{fig:g2_mm_squeezer_results} (b), where we illustrate how the detection of the \gtwo-function can provide us directly with comprehensive knowledge about the underlying spectral mode structure of the analyzed state.

In conclusion we have shown, that by measuring the second-order correlation function \(g^{(2)}\) of a multimode broadband twin-beam state one can probe the corresponding  distribution of spectral modes \(\lambda_k\). Our method displays the advantage that  correlation functions can be measured in a very practical way \cite{eckstein_highly_2011}, resulting in an approach that is much easier than realizing homodyne measurements, which require addressing individual modes. As a side remark we would like to point out that one can also determine the effective number of squeezers from the higher moments \(g^{(n)}\, , n \ge 2\) similar to the presented approach, yet \(g^{(2)}\) is already sufficient for our purposes.

\subsection{Probing the optical gain \(B\) of a multimode twin-beam squeezer via \(g^{(1,1)}\) measurements}
\begin{figure}[htb]
    \begin{center}
        \includegraphics[width=0.5\linewidth]{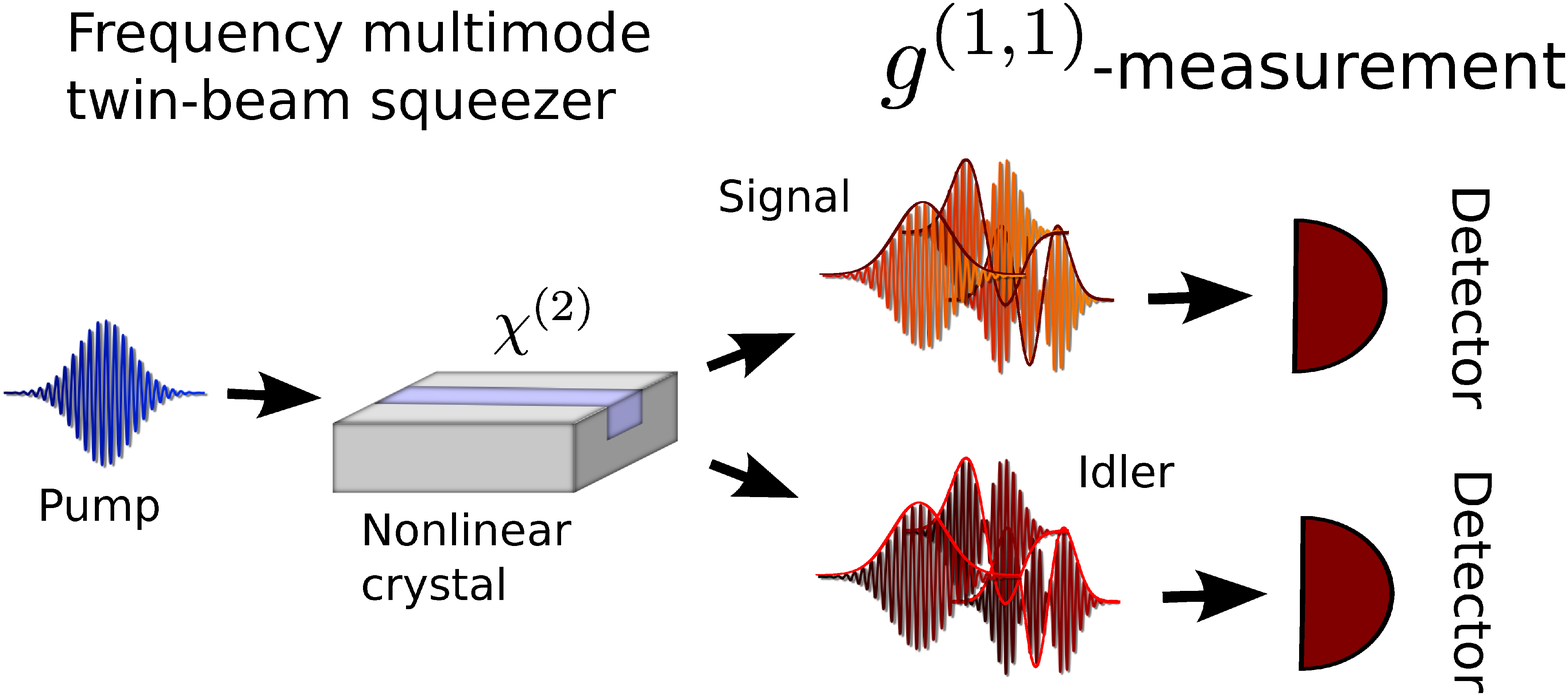}
    \end{center}
    \caption{Schematic setup to measure \(g^{(1,1)}\) of a multimode twin-beam squeezer generated via PDC.}
    \label{fig:g11_mm_squeezer_setup}
\end{figure}

\begin{figure}[htpb]
    \begin{center}
        \includegraphics[width=\linewidth]{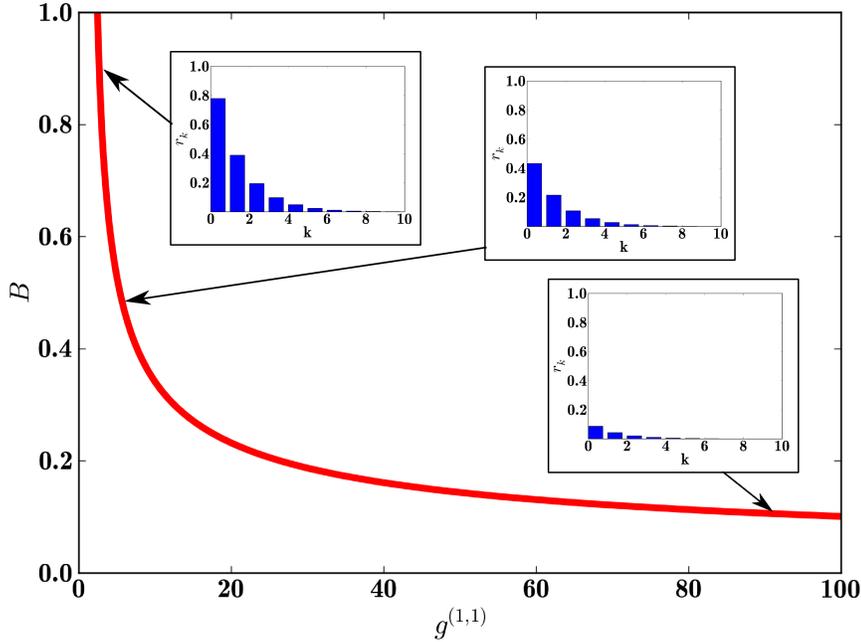}
    \end{center}
    \caption{The optical gain \(B\) plotted as a function of \goneone. For small values of \(B\) the correlation function \(g^{(1,1)}\) takes on a high value, yet rapidly decreases when the high gain regime is approached.}
    \label{fig:two_mode_squeezer_g11_analytic_inverse}
\end{figure}

In section \ref{sec:probing_twin_beam_squeezing_mode_distribution} we determined the number of modes in a loss resilient way by measuring \(g^{(2)}\) for low gains \(B\). Here we investigate the amount of the generated squeezing determined by the overall optical gain \(B\). In order to probe this value the setup has to be changed to measure the correlation function \(g^{(1,1)}\) of the generated twin-beam squeezer as presented in figure \ref{fig:g11_mm_squeezer_setup}.
Using  equation \eref{eq:broadband_cross-correlation_function} and \eref{eq:two_mode_squeezer_input_output_relation} we obtain for \(g^{(1,1)}\) the form
\begin{eqnarray}
    \nonumber
     \fl g^{(1,1)} = \frac{\sum_{k,l} \sinh^2(r_k) \sinh^2(r_l) + \sum_{k} \sinh^2(r_k) \cosh^2(r_k)}{\left[\sum_k \sinh^2(r_k)\right]^2} \\
    \fl \qquad = 1 + \underbrace{\frac{1}{\sum_k \sinh^2(r_k)}}_{1/\left<n\right>} + \underbrace{\frac{\sum_k \sinh^4(r_k)}{\left[\sum_k \sinh^2(r_k)\right]^2}}_{g^{(2)}-1}.
\label{eq:g11_two_mode_squeezer}
\end{eqnarray}
The relevant characteristics we exploit from this measurement is its dependence on both, the number of modes in the system, as given by the \gtwo-function \textit{and} the mean photon number in each arm, which is closely connected to the coupling coefficient \(B\). In the low gain regime (\(\sinh(r_k) \approx r_k\)), \(g^{(1,1)}\) simplifies to
\begin{eqnarray}
    g^{(1,1)} &\approx 1 + \frac{1}{B^2} + \underbrace{\frac{\sum_k \lambda_k^4}{\left[\sum_k \lambda_k^2 \right]^2}}_{g^{(2)}-1} 
    \approx \underbrace{g^{(2)}}_{\le 2} + \underbrace{\frac{1}{B^2}}_{\gg 1} 
     \approx  \frac{1}{B^2}. 
    \label{eq:g11_two_mode_squeezer_low_gain}
\end{eqnarray}
Hence, the optical gain is --- in the low gain regime --- obtained from the \goneone-measurement via the simple relation \(B \approx 1 / \sqrt{g^{(1,1)}}\). Mode dependencies of the coupling value \(B\) only occur at high squeezing strengths, where the relation diverges from equation \eref{eq:g11_two_mode_squeezer_low_gain} and takes on a more complicated form. In figure \ref{fig:two_mode_squeezer_g11_analytic_inverse} we plot the dependence of the overall coupling value \(B\) on \goneone --- as presented in equation \eref{eq:g11_two_mode_squeezer} --- which takes on a high value for small optical gains \(B\) but rapidly decreases when the high gain regime is approached.

In total measuring \(g^{(1,1)}\) gives direct \textit{loss-independent} access to the optical gain \(B\). This enables a loss tolerant probing of the generated mean photon number which, in the low gain regime, is even independent of the underlying mode structure.

Taking into account the prior knowledge we gained from section \ref{sec:probing_twin_beam_squeezing_mode_distribution}, we can now ascertain all parameters needed to fully determine the highly complex multimode state. The optical gain \(B\) defines not only the photon distribution, but quantifies the generated twin-beam squeezing, i.e. the available CV-entanglement in each mode. Note that all modes exhibit different entanglement parameters. Depending on the state and its respective mode distribution determined by the \gtwo-measurement all the entanglement could be generated in a single spectral mode where it is readily available for quantum information experiments or in a multitude of different squeezed modes. Note however, that after the state generation process multiple squeezers cannot be combined into a single optical mode by using only Gaussian operations, since this operation would be equivalent to continuous-variable entanglement distillation \cite{eisert_distilling_2002, fiurascaronek_gaussian_2002, giedke_characterization_2002}.

\section{Outlook}
In this paper we focused on the state characterization of ultrafast twin-beam squeezers in the time domain and their experimental analysis. The presented approach however is not limited to twin-beam squeezers:

On the one hand, our measurement technique also applies to probe the squeezing of ultrafast multimode \textit{single}-beam squeezers as presented in \ref{app:single_beam_squeezer}. On the other hand, our approach is easily adapted to spatial multimode squeezed states \cite{treps_quantum_2005, chalopin_multimode_2010, lassen_generation_2006}. These are characterized by measuring correlation functions that are broadband in the spatial domain, in a direct analogy to the spectral degree of freedom analyzed in this work.

\section{Conclusion}
We elaborated on the generation of multimode squeezed beams and their characterization with multimode broadband correlation functions. We expanded the formalism of correlation functions by including the effects of finite time resolution. These extended correlation function measurements serve as a versatile tool for the characterization of optical quantum states such as twin-beam squeezers. They provide a simple, straightforward and \textit{loss independent} way to investigate the characteristics of multimode squeezed states. Our findings are important for the field of efficient quantum state characterization and have already proven to be a useful experimental tool in the laboratory \cite{laiho_testing_2010, eckstein_highly_2011}.

\section{Acknowledgments}
This work was supported by the EC under the grant agreements CORNER (FP7-ICT-213681),
and QUESSENCE (248095).
Kati\'{u}scia N. Cassemiro acknowledges support from the Alexander von Humboldt foundation. 
The authors thank Agata M. Bra\'nczyk, Malte Avenhaus and Benjamin Brecht for useful discussions and helpful comments. \\
\bibliography{paper}

\begin{appendix}

\section{Multimode twin-beam squeezer generation via nonlinear optical processes}\label{app:multimode_two_mode_squeezer_generation}
\subsection{Generation of multimode twin-beam squeezers via parametric downconversion}\label{app:multimode_two_mode_squeezer_generation_PDC}
In the process of parametric downconversion squeezed states are generated by the interaction of a strong pump field with the \(\chi^{(2)}\)-nonlinearity of a crystal. Regarding the generation of twin-beam squeezers the Hamiltonian of the corresponding three-wave-mixing process is given by \cite{mauerer_how_2009, braczyk_optimized_2010, grice_spectral_1997}:
\begin{eqnarray}
    \hat{H}_{PDC} = \int_{-\frac{L}{2}}^{\frac{L}{2}} \mathrm d z \, \chi^{(2)}  \hat{E}^{(+)}_p(z,t) \hat{E}^{(-)}_s(z,t) \hat{E}_i^{(-)}(z,t) + h.c.
    \label{eq:pdc_twin-beam_hamiltonian}
\end{eqnarray}
where we focused on a collinear interaction of all three beams. In equation \eref{eq:pdc_twin-beam_hamiltonian} \(L\) labels the length of the medium, \( \chi^{(2)} \) the nonlinearity of the crystal, and \(\hat{E}_p^{(+)}(z,t)\), \(\hat{E}^{(-)}_s(z,t)\), \(\hat{E}^{(-)}_i(z,t)\) the pump, the signal and the idler fields. The electric field operators used in  equation \eref{eq:pdc_twin-beam_hamiltonian} are defined as follows
\begin{eqnarray}
    \fl \qquad \hat{E}_{x}^{(-)}(z,t) =  \hat{E}_{x}^{(+)\dagger}(z,t) =  C \int \mathrm d \omega_{x} \, \exp\left[-\imath\left(k_{x}(\omega)z + \omega t\right)\right] \hat{a}^\dagger_{x}(\omega)  \label{eq:electric_field_operator},
\end{eqnarray}
in which we have merged all constants and slowly varying field amplitudes in the overall parameter \(C\). In order to simplify the Hamiltonian we treat the strong pump field as a classical wave
\begin{eqnarray}
    \hat{E}_p^{(+)}(z,t) \Rightarrow E_p(z,t) =  \int \mathrm d \omega_p \, \alpha(\omega_p) \exp\left[\imath\left(k_p(\omega_p)z + \omega_p t\right)\right].
    \label{eq:electric_pump_field}
\end{eqnarray}
Here \(\alpha(\omega_p) = A_p \exp\left[(\omega_p - \mu_p)^2/ (2\sigma_p^2)\right]\) is the Gaussian pump envelope function generated by an ultrafast laser system, consisting of a field amplitude \(A_p\), a central pump frequency \(\mu_p\), and a pump width \(\sigma_p\).

The PDC Hamiltonian in equation \eref{eq:pdc_twin-beam_hamiltonian} generates the following unitary transformation:
\begin{eqnarray}
    \hat{U} = \exp\left[-\frac{\imath}{\hbar} \int_{-\infty}^{\infty}\mathrm d t' \, \hat{H}_{PDC}(t')\right]
    \label{eq:unitary_operator_two_mode_squeezer}
\end{eqnarray}
In the low downconversion regime we can ignore the time-ordering of the electric field operators  \cite{wasilewski_pulsed_2006, lvovsky_decomposing_2007} and directly evaluate the time integration. This yields a delta-function \(2 \pi \delta(\omega_s + \omega_i - \omega_p)\) and hence allows us to perform the integral over the pump frequency \(\omega_p\). Equation \ref{eq:unitary_operator_two_mode_squeezer} can be re-expressed as
\begin{eqnarray}
    \nonumber
    \fl \hat{U} = \exp \left[ -\frac{\imath}{\hbar} \left(A' \int_{-\frac{L}{2}}^{\frac{L}{2}} \mathrm d z \int \mathrm d \omega_s \int \mathrm d \omega_i \,  \right. \right. \\
    \left. \left. \times \alpha(\omega_s + \omega_i) \exp\left[\imath \Delta k z\right]  \hat{a}_s^\dagger(\omega_s) \hat{a}_i^\dagger(\omega_i) + h.c. \right) \right],
\end{eqnarray}
in which \(\Delta k = k_p(\omega_s + \omega_i) - k_s(\omega_s) - k_i(\omega_i)\) is the so called phase-mismatch and \(A'\) accumulates all constants. Finally, we perform the integration over the length of the crystal and obtain 
\begin{eqnarray} 
    \fl \qquad \hat{U} = \exp\left[ -\frac{\imath}{\hbar} \left( A \int \mathrm d \omega_s \int \mathrm d \omega_i \, \alpha(\omega_s + \omega_i)
     \phi(\omega_s, \omega_i) \hat{a}_s^\dagger(\omega_s) \hat{a}_i^\dagger(\omega_i) + h.c. \right)\right],
    \label{eq:unitary_typeII_process_derivation}
\end{eqnarray}
where \(\phi(\omega_s, \omega_i) = \mathrm{sinc}\left(\frac{\Delta k L}{2}\right)\) is referred to as the phasematching function.
The latter combined with the pump distribution \(\alpha(\omega_s + \omega_i)\) gives the overall frequency distribution or joint spectral amplitude \(f(\omega_s, \omega_i)\) of the generated state. The final unitary squeezing operator of the downconversion process is
\begin{eqnarray}
    \hat{U} = \exp\left[-\frac{\imath}{\hbar}\underbrace{\left( A \int \mathrm d \omega_s \int \mathrm d \omega_i f(\omega_s, \omega_i) \hat{a}_s^\dagger(\omega_s) \hat{a}_i^\dagger(\omega_i) + h.c. \right)}_{\hat{H}_{eff}}  \right].
    \label{eq:unitary_PDC_twin_beam_derivation}
\end{eqnarray}

The \(\mathrm{sinc}\) function appearing in equation \ref{eq:unitary_PDC_twin_beam_derivation} can be approximated by a Gaussian distribution
\begin{eqnarray}
    \fl \qquad \phi(\omega_s, \omega_i) =  \mathrm{sinc}\left(\frac{\Delta k(\omega_s,\omega_i) L}{2}\right) 
                            \approx \exp\left[-0.193 \left(\frac{\Delta k(\omega_s, \omega_i) L}{2}\right)^2\right].
    \label{eq:phasematching_function_approximation}
\end{eqnarray}
With this simplification the joint frequency distribution \(f(\omega_s, \omega_i)\) takes on the form of a two-dimensional Gaussian distribution. Applying this approximation the exact squeezer distribution is accessible as presented in section \ref{sec:probing_twin_beam_squeezing}. 

\subsection{Generation of multimode twin-beam squeezers via four-wave-mixing}
In a four-wave-mixing (FWM) process two strong pump fields interact with the \(\chi^{(3)}\)-nonlinearity of a fiber to create two new electric fields. If the two generated fields are distinguishable the Hamiltonian of the process is given by \cite{chen_quantum_2007}
\begin{eqnarray}
    \fl \qquad \hat{H}_{\mathrm{FWM}} = \int_{-\frac{L}{2}}^{\frac{L}{2}} \mathrm d z \, \chi^{(3)}  \hat{E}^{(+)}_{p1}(z,t) \hat{E}^{(+)}_{p2}(z,t) \hat{E}^{(-)}_s(z,t) \hat{E}_i^{(-)}(z,t) + h.c. \,\, .
    \label{eq:fwm_twin-beam_hamiltonian}
\end{eqnarray}
Again, we assume a collinear interaction of all interacting beams. The electric fields for signal, idler and pump are defined in equations \eref{eq:electric_field_operator} and \eref{eq:electric_pump_field}. Performing the same steps as in \ref{app:multimode_two_mode_squeezer_generation_PDC} we obtain a similar unitary transformation
\begin{eqnarray}
    \fl \qquad \hat{U} = \exp\left[ -\frac{\imath}{\hbar} \underbrace{\left( A \int \mathrm d \omega_s \int \mathrm d \omega_i \, f_{\mathrm{FWM}}(\omega_s, \omega_i) \hat{a}_s^\dagger(\omega_s) \hat{a}_i^\dagger(\omega_i) + h.c. \right)}_{\hat{H}_{eff}}\right] .
    \label{eq:unitary_FWM_twin_beam_derivation}
\end{eqnarray}
Equation \eref{eq:unitary_FWM_twin_beam_derivation} resembles equation \eref{eq:unitary_PDC_twin_beam_derivation} with the exception of the joint frequency distribution \(f_{\mathrm{FWM}}(\omega_s, \omega_i)\) which takes on a more complicated shape in comparison to the PDC case
\begin{eqnarray}
    \fl \qquad f_{\mathrm{FWM}}(\omega_s, \omega_i) = \int \mathrm d \omega_p \,  \alpha(\omega_{p}) \alpha(\omega_s + \omega_i - \omega_p)\, \mathrm{sinc}\left( \frac{\Delta k (\omega_p, \omega_s, \omega_i) L}{2} \right) .
\end{eqnarray}
By comparing the unitary transformation in equation \eref{eq:unitary_typeII_process_derivation} and \eref{eq:unitary_FWM_twin_beam_derivation} it is apparent that the two different processes both create the same fundamental quantum state: Multimode twin-beam squeezers.

\section{Multimode single-beam squeezers}\label{app:single_beam_squeezer}
In the main body of the paper we discussed the characterization of multimode twin-beam squeezers. Here we call attention to the fact that the broadband multimode correlation function formalism is also applicable to probe multimode single-beam squeezed states.

\subsection{Generation of multimode single-beam squeezers}\label{app:single_beam_squeezer_generation}
Single-beam squeezers are created by PDC and FWM processes similar to the twin-beam states. The difference between the twin-beam and single-beam squeezer generation is that in the latter the generated beams are emitted into the same optical mode, whereas in the former two different optical modes are generated as discussed in \ref{app:multimode_two_mode_squeezer_generation}.

The PDC Hamiltonian generating a single-beam squeezer is given by
\begin{eqnarray}
    \hat{H} = \int_{-\frac{L}{2}}^{\frac{L}{2}} \mathrm d z \, \chi^{(2)}  \hat{E}^{(+)}_p(z,t) \hat{E}^{(-)}(z,t) \hat{E}^{(-)}(z,t) + h.c. \,\, .
    \label{eq:pdc_single-beam_hamiltonian}
\end{eqnarray}
Performing the same steps as in the case of twin-beam generation we obtain the unitary transformation
\begin{eqnarray}
    \hat{U} = \exp\left[-\frac{\imath}{\hbar}\underbrace{\left( A \int \mathrm d \omega_s \int \mathrm d \omega_i f(\omega_s, \omega_i) \hat{a}^\dagger(\omega_s) \hat{a}^\dagger(\omega_i) + h.c. \right)}_{\hat{H}_{eff}}  \right].
\end{eqnarray}
If the joint spectral distribution \(f(\omega_s, \omega_i)\) is engineered to be symmetric under permutation of signal and idler, the Schmidt decomposition is given by:
\begin{eqnarray}
    -\frac{\imath}{\hbar} A f(\omega_s, \omega_i) = \sum_k r_k \phi_k^*(\omega_s) \phi_k^*(\omega_i) \,\,\, \mathrm{and}\\
    -\frac{\imath}{\hbar} A^* f^*(\omega_s, \omega_i) = -\sum_k r_k \phi_k(\omega_s) \phi_k(\omega_i)
    \label{eq:singular_value_decomposition_single_beam}
\end{eqnarray}
Introducing broadband modes we obtain the multimode broadband unitary transformation
\begin{eqnarray}
    \nonumber
    \hat{U} &= \exp\left[\sum_k r_k \hat{A}_k^\dagger \hat{A}_k^\dagger - h.c.\right] \\
    \nonumber
            &= \bigotimes_k \exp\left[r_k \hat{A}_k^\dagger \hat{A}_k^\dagger - h.c.\right]\\
            &= \bigotimes_k \hat{S}(-r_k).
    \label{eq:OPA_unitary_single_beam}
\end{eqnarray}
This is exactly the form of a frequency multimode single-beam squeezed state \cite{barnett_methods_2003}.
Or written in the Heisenberg picture:
\begin{eqnarray}
    \hat{A}_k = \cosh(r_k) \hat{A}_k + \sinh(r_k) \hat{A}_k^\dagger
\end{eqnarray}
Single-beam squeezers are --- like twin-beam squeezers --- widely employed in quantum optics experiments \cite{zhu_photocount_1990, sasaki_multimode_2006}. As in the twin-beam squeezer case the same states are generated by properly engineered FWM processes.

\subsection{Probing frequency multimode single-beam squeezers via correlation function measurements}
\begin{figure}[htb]
    \begin{center}
        \includegraphics[width=0.9\linewidth]{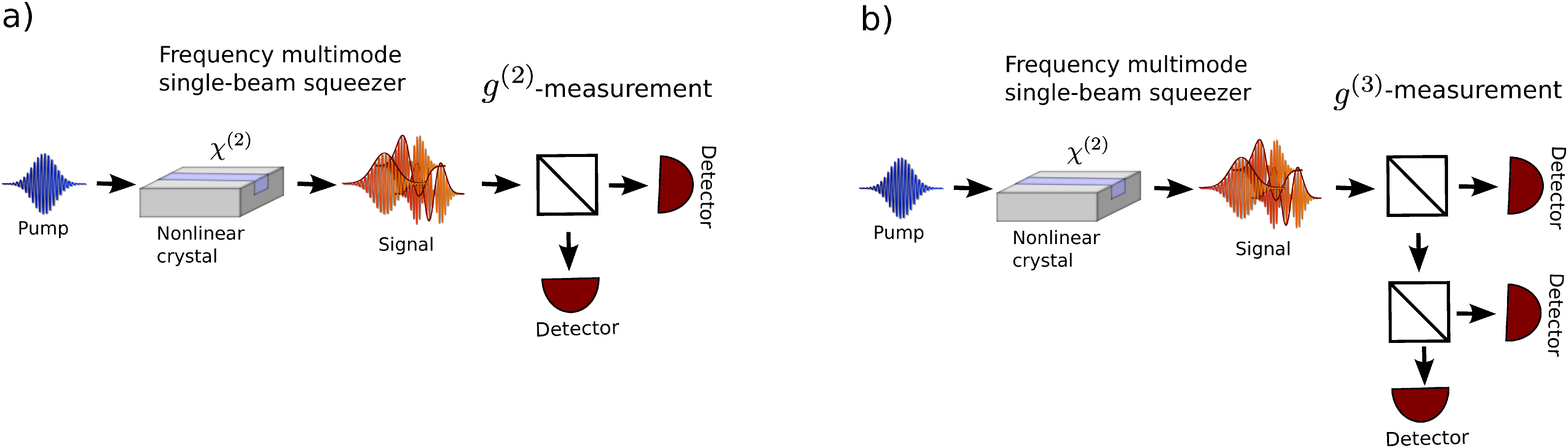}
    \end{center}
    \caption{Schematic setup to measure a) \(g^{(2)}\) and b) \(g^{(3)}\) of a frequency multimode single-beam squeezer.}
    \label{fig:squeezer_two_mode_g2_g3_setup}
\end{figure}

\begin{figure}[htb]
    \begin{center}
        \includegraphics[width=\linewidth]{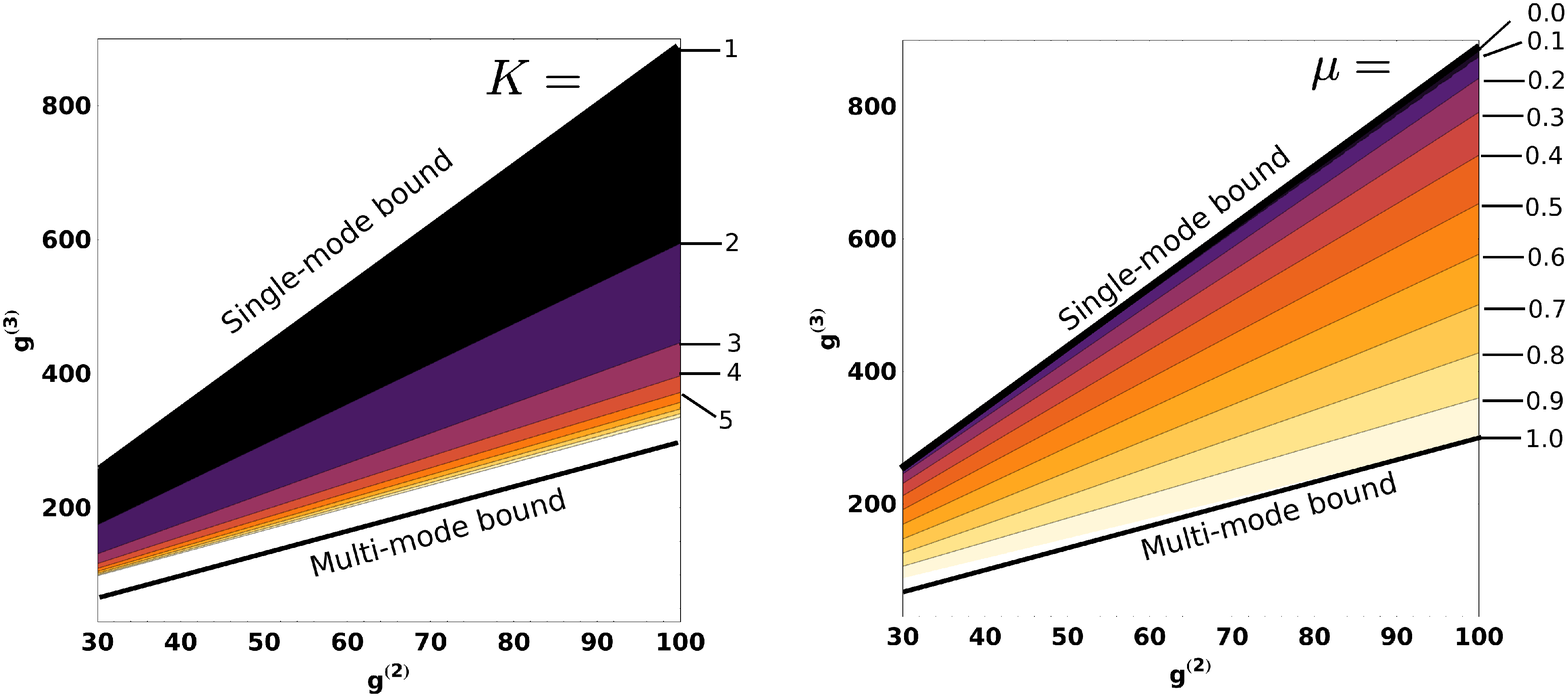}
    \end{center}
    \caption{\(g^{(3)}\) as a function of \(g^{(2)}\) for various multimode single-beam squeezers. The effective number of modes and the thermal mode distributions parameter \(\mu\) of a multimode single-beam squeezer are encoded in the slope.}
    \label{fig:g3_vs_g2_mm_squeezer_K_mu_dependence}
\end{figure}

\begin{figure}[htpb]
    \begin{center}
        \includegraphics[width=\linewidth]{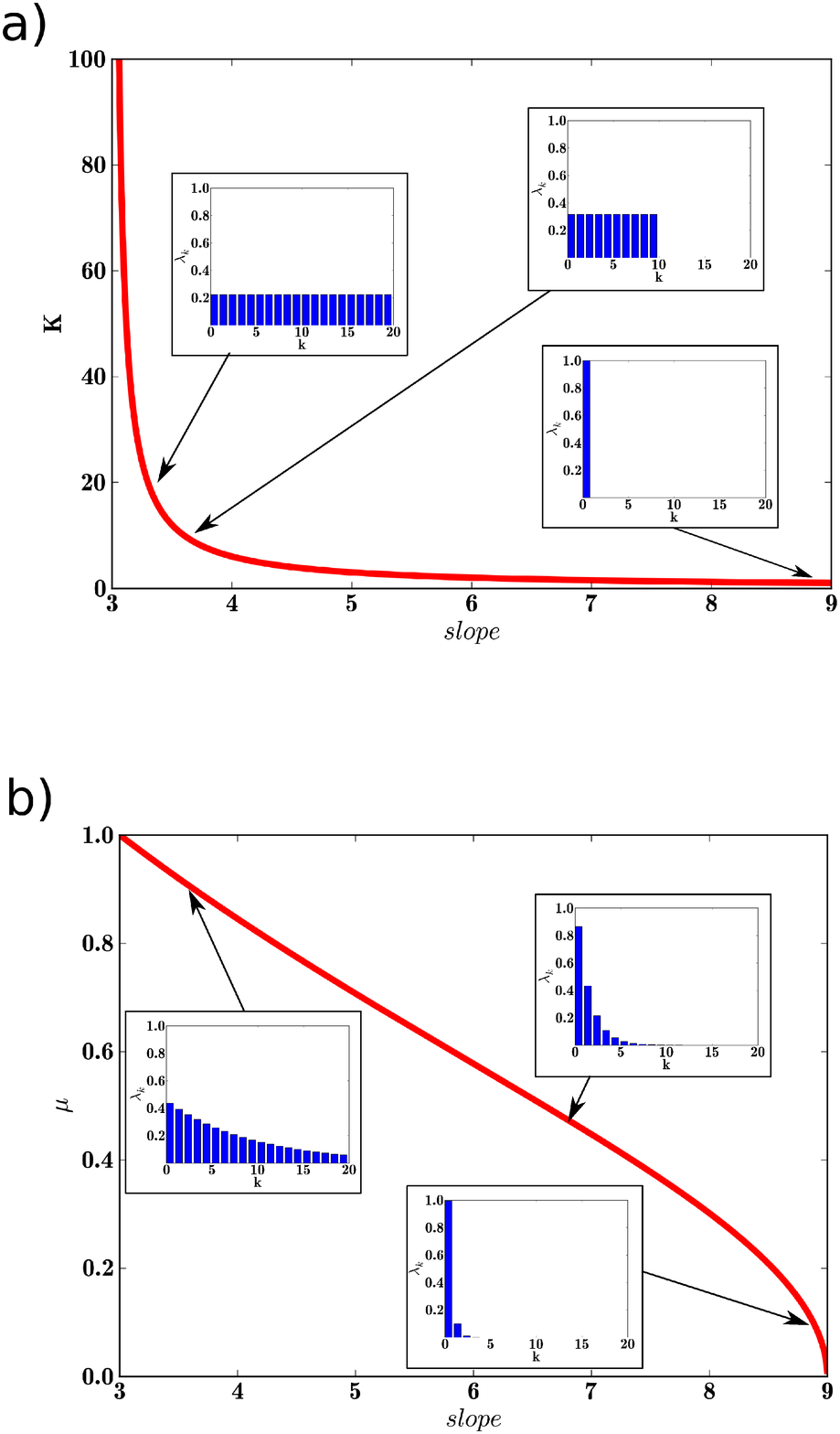}
    \end{center}
    \caption{a) Effective mode number \(K\) as a function of the slope of \(g^{(3)}[g^{(2)}]\). b) Thermal mode distribution \(\mu\) as a function of the slope of \(g^{(3)}[g^{(2)}]\) for multimode single-beam squeezed states.}
    \label{fig:g3_vs_g2_mm_squeezer_K_mu_analysis}
\end{figure}

In order to characterize the generated states we have to determine the optical gain \(B\) and mode distribution \(\lambda_k\) as in the case of multimode twin-beam squeezers (see section \ref{sec:probing_twin_beam_squeezing}). Therefore, we adapt the scheme presented in section \ref{sec:probing_twin_beam_squeezing} and probe the correlation functions \(g^{(2)}\) and \(g^{(3)}\) as sketched in figure \ref{fig:squeezer_two_mode_g2_g3_setup}. 
For a multimode single-beam squeezer they can be written as: 
\begin{eqnarray}
    g^{(2)} =& 1 + 2 \frac{\sum_k \sinh^4(r_k)}{\left[\sum_k \sinh^2(r_k)\right]^2} + \underbrace{\frac{1}{\sum_k \sinh^2(r_k)}}_{1/\left<n\right>} \,\,\,\,\, \mathrm{and}\\
    \nonumber
    g^{(3)} =& 1 + 6 \frac{\sum_k \sinh^4(r_k)}{\left[\sum_k \sinh^2(r_k)\right]^2} + 8 \frac{\sum_k \sinh^6(r_k)}{\left[\sum_k \sinh^2(r_k)\right]^3}\\
    & + \frac{3}{\sum_k \sinh^2(r_k)} + 6 \frac{\sum_k \sinh^4(r_k)}{\left[\sum_k \sinh^2(r_k)\right]^3}.
    \label{eq:mm_sm_squeezer_correlations}
\end{eqnarray}
In the single-beam case however \(g^{(2)}\) does not directly yield the effective number of modes \(K\) or thermal mode distribution parameter \(\mu\) as for the multimode twin-beam squeezers in equation \eref{eq:gtwo_mm_two_mode_squeezer}. A joint measurement of \(g^{(2)}\) and \(g^{(3)}\) is necessary, as sketched in figure \ref{fig:g3_vs_g2_mm_squeezer_K_mu_dependence}. 
Clearly the effective mode number \(K\) and the thermal mode distribution \(\mu\) are given by the slope \(s\) of \(g^{(3)}\) vs. \gtwo. In figure \ref{fig:g3_vs_g2_mm_squeezer_K_mu_analysis} we plotted the explicit dependence of \(K\) and \(\mu\) on the slope \(s\). Surprisingly the functions exhibit almost the same shape as in the twin-beam squeezer case.

\begin{figure}[htpb]
    \begin{center}
        \includegraphics[width=\linewidth]{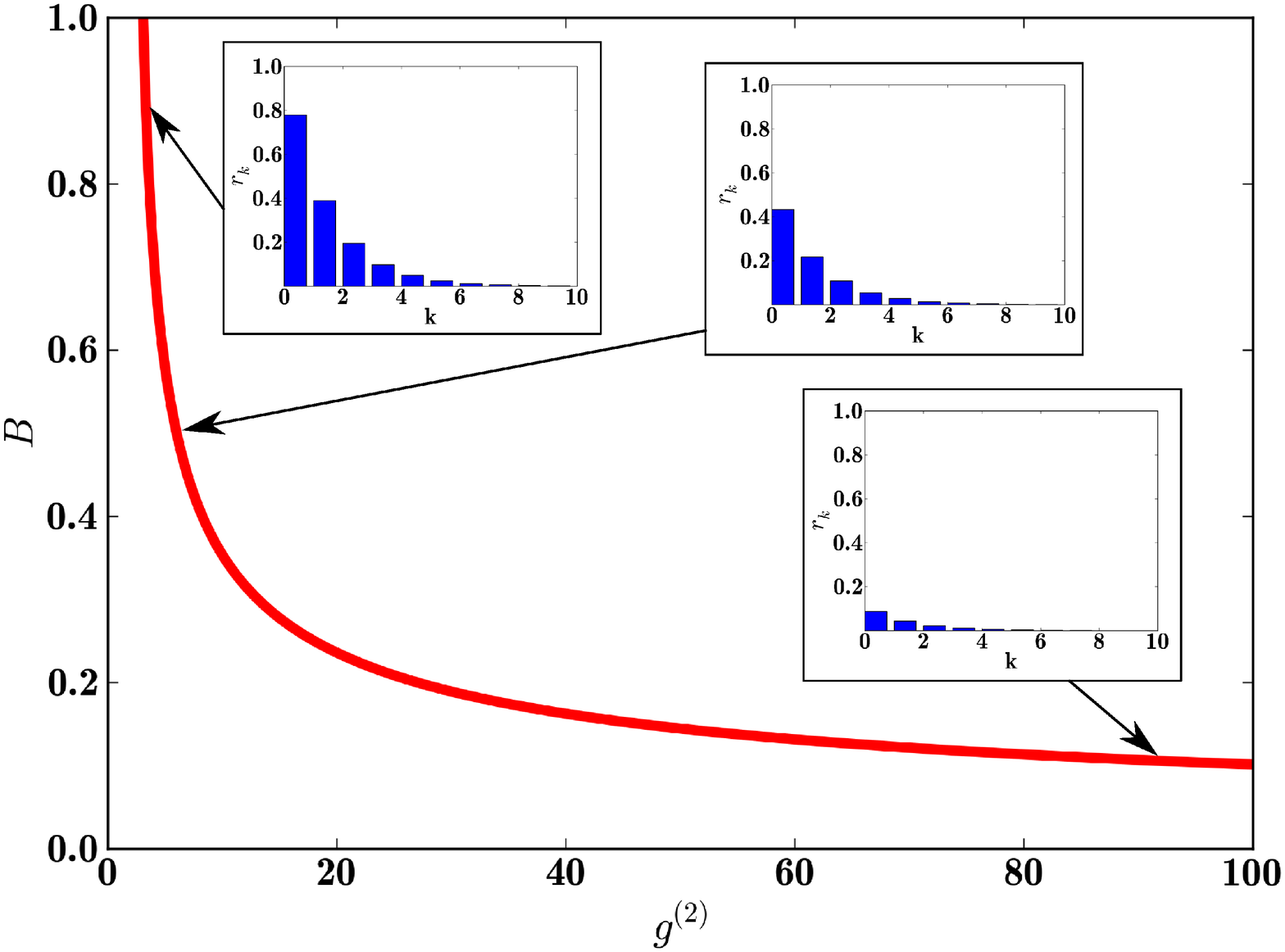}
    \end{center}
    \caption{Optical gain \(B\) as a function of \(g^{(2)}\) for a multimode single-beam squeezed state.}
    \label{fig:g3_vs_g2_mm_squeezer_B}
\end{figure}

In order to obtain the gain of a multimode single-beam squeezer a single \gtwo-measurement is sufficient which is sensitive towards the coupling value \(B\) as presented in figure \ref{fig:g3_vs_g2_mm_squeezer_B} (similar to the \goneone-measurement in the twin-beam squeezer case). In the low gain regime it is given via the relation \(B = 1 / \sqrt{g^{(2)}}\).  Again, while describing a different system, the shape of the function \(B[g^{(2)}]\) is very similar to the twin-beam squeezer case.

In total the theoretical description and derivation of multimode single-beam squeezers is very similar to the mathematics behind multimode twin-beam states. These similarities translate to multimode correlations functions which are able to probe the generated optical gain \(B\) and mode distribution \(\lambda_k\) as in the twin-beam case.
\end{appendix}
\end{document}